\documentclass{emulateapj}
\usepackage{natbib}
\usepackage{graphicx}

\newcommand{\gammabar}{\bar{\gamma}}
\newcommand{\eqb}{\begin{eqnarray}}
\newcommand{\eqe}{\end{eqnarray}}
\newcommand{\diff}{\textrm{d}}

\slugcomment{To appear in ApJ Letters}
\shorttitle{Radiative signatures of relativistic shocks}
\begin{document}

\title{Radiative Signatures of Relativistic Shocks}

\author{John G. Kirk and Brian Reville}
\affil{Max-Planck-Institut f\"ur Kernphysik, Postfach 10~39~80,
69029 Heidelberg, Germany}
\email{john.kirk@mpi-hd.mpg.de}

\begin{abstract}
  Particle-in-cell simulations of relativistic, weakly magnetized
  collisionless shocks show that particles can gain energy by
  repeatedly crossing the shock front.  This requires scattering off
  self-generated small length-scale magnetic fluctuations.  The
  radiative signature of this first-order Fermi acceleration mechanism
  is important for models of both the prompt and afterglow emission in
  gamma-ray bursts and depends on the strength parameter $a=\lambda
  e|\delta B|/mc^2$ of the fluctuations ($\lambda$ is the length-scale
  and $|\delta B|$ the magnitude of the fluctuations.)  For electrons
  (and positrons), acceleration saturates when the radiative losses
  produced by the scattering cannot be compensated by the energy
  gained on crossing the shock.  We show that 
  this sets an upper limit on both the electron Lorentz factor:
  $\gamma\,<\,10^6
  \left(n/1\,\textrm{cm}^{-3}\right)^{-1/6}{\bar{\gamma}}^{1/6}$ and
  on the energy of the photons radiated during the scattering process:
  $\hbar\omega_{\rm max}\,<\,40\,{\rm Max}(a,1) \left(n/1\,
    \textrm{cm}^{-3}\right)^{1/6}{\bar{\gamma}}^{-1/6}\,\textrm{eV}$,
  where $n$ is the number density of the plasma and ${\bar{\gamma}}$
  the thermal Lorentz factor of the downstream plasma, provided 
$a<a_{\rm crit}\sim10^6$. 
  This rules out \lq
  jitter\rq\ radiation on self-excited fluctuations with $a<1$ as a source of
  gamma-rays, although high-energy photons might still be produced
  when the jitter photons are upscattered in an analog of the
  synchrotron self-Compton process. 
In fluctuations with $a>1$,
  radiation is generated by the standard synchrotron mechanism, and the 
  maximum photon energy rises linearly with $a$, until saturating at
$70\,$MeV, when $a=a_{\rm crit}$. 
\end{abstract}

\keywords{radiation mechanisms: non-thermal --- acceleration of particles --- gamma rays: bursts}

\section{Introduction}

In astrophysics, the most widely discussed mechanism of particle
acceleration is the first-order Fermi process operating at
collisionless shocks. It is based on the idea that particles
undergo stochastic elastic scatterings both upstream and downstream of the
shock front. This causes particles
to wander across the shock repeatedly. On each crossing, they receive an
energy boost as a result of the relative motion of the upstream
and downstream plasmas. At non-relativistic shocks, scattering 
causes particles to diffuse in space, and the mechanism,
termed \lq diffusive shock acceleration\rq, is widely thought to be
responsible for the acceleration of cosmic rays in supernova
remnants. 
At relativistic shocks, the transport process is not 
spatial diffusion, but the first-order Fermi
mechanism operates nevertheless \cite[for reviews see][]{kirkduffy99, hillas05}.  
In fact, the first {\em ab initio}
demonstrations of this process using particle-in-cell (PIC)
simulations have recently been presented for the relativistic case
\citep{spitkovsky08b,martinsetal09,sironispitkovsky09a}.

Several factors, such as the lifetime of the shock front, or its
spatial extent, can limit the energy to which particles can be
accelerated in this process. However, even in the absence of these, 
acceleration will ultimately cease when the radiative
energy losses that are inevitably associated with the scattering
process overwhelm the energy gains obtained upon crossing the shock.
Exactly when this happens depends on the details of the scattering
process.

In the non-relativistic case, the diffusion coefficient is frequently
parameterized in terms of its lower limit, known as Bohm diffusion, at
which the particle mean free path equals the gyro radius. In this
case, particles that achieve the highest possible energy radiate
synchrotron photons up to an energy 
$\sim150\eta\left(v_{\rm s}/c\right)^2\,$MeV,
where $v_{\rm s}$ is the speed of the shock
front and $\eta$($\le1$) is the inverse ratio of the diffusion
coefficient to its value in the Bohm limit.  Interestingly, this
photon energy is independent of the magnetic field strength.

In PIC simulations of weakly magnetized shocks, however,
particles appear to undergo small-angle scatterings on fluctuations in
the electromagnetic fields 
that are driven by the Weibel instability. These have a
length scale roughly equal to the
inertial length $c/\omega_{\rm p}$, where 
$\omega_{\rm p}=\sqrt{4\pi n e^2/\gammabar m}$ 
is the relativistic electron plasma frequency, and
$\gammabar$ is the Lorentz factor of the average thermal motion.
As we show below (see Eq. (\ref{nscatters}))
the resulting mean free path of a particle of Lorentz factor $\gamma$ 
is proportional to $\gamma^2$, 
rather than the linear dependence expected in Bohm diffusion.
This modifies the maximum energy of the synchrotron photons 
\citep{derishev07}, but the radiation emitted when a particle is 
deflected through very small angles differs 
significantly from synchrotron radiation
\citep{landaulifshitz},
and can, in principle, produce more energetic photons.
This has led to the suggestion that \lq\lq jitter\rq\rq\ radiation 
from shock-accelerated particles
is responsible for both the prompt emission \citep{medvedev06}
and the afterglow \citep{medvedevetal07} from gamma-ray bursts.
In this {\em Letter} we show that the inherent 
weakness of the scattering produced by Weibel-driven turbulence
implies that radiation losses quench first-order Fermi acceleration
relatively quickly. In this case, the 
maximum photon energy is given by (\ref{freqconstraint})
and particles are unable to produce X-ray or 
gamma-ray photons in the downstream rest frame. 
 
\section{Relativistic shock fronts}

Recent particle-in-cell simulations of relativistic shocks have
revealed fundamental differences between the magnetized and
unmagnetized cases.  The most detailed results are available for
electron-positron pair plasmas, which are likely to be found in
gamma-ray bursts and pulsar winds.  
The upstream and downstream plasmas can be characterized by 
magnetic field strengths $B_{\rm u,d}$ and number densities
$n_{\rm u,d}$ measured in the respective rest frames. The upstream
plasma is cold, and streams into the shock with a Lorentz factor
$\gammabar$, roughly equal to the thermal Lorentz factor of the downstream
plasma. The magnetization parameters in each region can be written:
\eqb
\begin{array}{l}
\sigma_{\rm u}=B_{\rm u}^2/(4\pi n_{\rm u} mc^2)
\\
\sigma_{\rm d}=B_{\rm d}^2/(4\pi \gammabar n_{\rm d} mc^2)
\end{array}
\eqe
At a parallel shock, $B_{\rm u}=B_{\rm d}$ and 
$\sigma_{\rm d}\ll\sigma_{\rm u}$.
If the magnetic field is compressed at an oblique shock front,
one expects $\sigma_{\rm u}\sim\sigma_{\rm d}$, since $B_{\rm d}/B_{\rm u}\sim
n_{\rm d}/n_{\rm u}\sim3\gammabar$. However, if magnetic 
field is generated, then
$\sigma_{\rm d}\gg \sigma_{\rm u}$.  
It is found \citep{sironispitkovsky09a} that \lq
unmagnetized\rq\ shocks, with $\sigma_{\rm u}<10^{-3}$, are mediated by the
Weibel instability. This is true also for shocks with an upstream
magnetic field parallel to the shock normal, provided $\sigma_{\rm u}<0.1$. On
the other hand, more strongly magnetized or oblique shocks are
mediated by the synchrotron maser instability that operates when
particles streaming into the shock start to gyrate about the
compressed downstream magnetic field.  In each case, it is possible to
estimate a length scale on which turbulence is initially
generated in the electromagnetic field. For Weibel-mediated
(unmagnetized) shocks, this is 
\eqb
\lambda_{\rm w}&=&\ell_{\rm w} c/\omega_{\rm p}
\label{weibellength}
\eqe
with $\ell_{\rm w}\sim 10$, according to \citet{sironispitkovsky09a}. Here, 
$\omega_{\rm p}$
is the local plasma frequency. Because 
$n_{\rm d}/n_{\rm u}\sim\gammabar$, the plasma frequencies in the upstream and downstream
media are roughly equal,
and so the wavelength of the turbulence generated is 
approximately the same.
For the purpose of estimating radiative signatures of accelerated
particles, it is more convenient to characterize the fluctuations
in terms of their \lq strength\rq, or 
\lq wiggler\rq\ parameter $a$, defined as the 
ratio of their length scale to the length defined by the 
turbulent field strength: $ a=\lambda e|\delta B|/mc^2$.
In the upstream and downstream media (subscripts \lq\lq u\rq\rq\ and 
\lq\lq d\rq\rq) one has
\eqb
\begin{array}{l}
a_{\rm w,u}
\approx\ell_{\rm w}\sigma_{\rm u}^{1/2}
\\
a_{\rm w,d}\approx\ell_{\rm w}\gammabar\sigma_{\rm d}^{1/2}
\end{array}
\label{weibelstrength}
\eqe
This suggests that for Weibel-mediated shocks in pair plasmas 
the strength parameters are likely to be small ($a\lesssim1$), 
although moderate values ($a\gtrsim1$) are possible
if the self-generated field grows locally to levels close to 
equipartition.
Carrying out the same analysis for unmagnetized electron-ion shocks 
results in an additional factor of ($m_{\rm i}/m_{\rm e}$) in the value
of the strength parameter, since the dominant length-scale in this case is
the ion skin-depth \citep{spitkovsky08a}. Provided $\sigma_{u,d}\ll 1$, 
moderate strength parameters can still be expected.

For magnetized shocks mediated by the synchrotron maser instability
\citep{lyubarsky07},
the characteristic length in the downstream plasma
$\lambda_{\rm s,d}$
is dictated by the requirement that  
the incoming particles be significantly deflected: 
\eqb
\lambda_{\rm s,d}&=&\ell_{\rm s}\gammabar m c^2/eB_{\rm d}
\eqe
with $\ell_{\rm s}\sim1$. This generates electromagnetic waves
of strength parameter
\eqb
a_{\rm s,d}&=&\ell_{\rm s}\gammabar
\label{synchstrength}
\eqe
These waves propagate into the upstream medium without change of their
(Lorentz invariant) strength parameter: $a_{\rm s,u}=a_{\rm s,d}$.

To summarize, the strength parameters associated with 
shock-generated turbulence are likely to be small ($a\lesssim1$) 
in the case of Weibel-mediated pair shocks, but moderate
$a\sim\gammabar$ 
in the case of electron-ion shocks and magnetized shocks mediated by the 
synchrotron instability. 

\section{Particle acceleration}

Fermi acceleration at relativistic shocks differs from that at 
nonrelativistic shocks in that it is sensitive to the 
physics of the particle scattering process
\citep[e.g.,][]{kirkduffy99}. 
Typically, it is assumed that the particles undergo stochastic
small-angle deflections through interactions with magnetic irregularities.
According to the strength parameter of these irregularities,
the transport of the particles can then 
be divided into two distinct regimes, which we call
\lq ballistic\rq\ and \lq helical\rq. 
In ballistic transport, the scattering
mean free path is shorter than the  
gyroradius in the local field, so that the curvature radius of
a trajectory is larger than the length scale on which this quantity 
fluctuates. This kind of transport is produced by fluctuations with $a<\gamma$.
On the other hand, in the helical transport regime,
gyro motion is only slightly perturbed. Particles have sufficient time to
gyrate about the field while their pitch angles and
guiding-center positions diffuse. This requires fluctuations with $a>\gamma$.
This has a strong influence on the acceleration process and 
its interplay with radiation losses.

\subsection{Ballistic transport regime}
At a relativistic shock, a particle remains in the upstream medium 
until it has
been deflected, on average, 
through an angle of $1/\gammabar$ in the upstream rest frame. 
A turbulent fluctuation of strength parameter $a$, 
deflects a particle of Lorentz
factor $\gamma$ through an angle $a/\gamma$. Provided
this angle is small, the diffusion coefficient is simply
$\mathcal{D}_\theta=a^2\nu_{\rm sc}/\gamma^2$, where $\nu_{\rm sc}$ is the mean
scattering frequency. The average number of scatterings in the
upstream medium between shock encounters is therefore
\eqb
N_{\rm scatt,u}&\approx& \left(\gamma/a_{\rm u}\gammabar\right)^2
\label{nscatters}
\eqe
At each scattering, the power radiated in photons
by the energetic particle can be estimated from Larmor's formula:
\eqb
\left.
\frac{\Delta\gamma}{\gamma}\right|_{\rm loss,u}
&\approx&\frac{2a_{\rm u}^2e^2\gamma}{3mc^2\lambda_{\rm u}}
\eqe
assuming inverse Compton losses can be neglected.
For kinematic
reasons, the average energy gain per cycle is roughly a
factor of two \citep{achterbergetal01},
so that the acceleration process will saturate when
the energy lost in the upstream medium is roughly 
$\gamma mc^2$. This implies
$N_{\rm scatt,u}\left.\Delta\gamma/\gamma\right|_{\rm loss,u}<1$, or
\eqb
\gamma&<&\left(\frac{3mc^2\lambda_{\rm u}\gammabar^2}{2e^2}\right)^{1/3}
\label{wglimitup}
\eqe

Applying the same argument, energy loss by scattering in the downstream
medium, where particles must be scattered through an angle of roughly $\pi/2$, 
imposes the condition
\eqb
\gamma&<&\left(\frac{3mc^2\lambda_{\rm d}}{2e^2}\right)^{1/3}
\label{wglimitdown}
\eqe
which, since $\lambda_{\rm d}\sim\lambda_{\rm u}$, is more restrictive.

The ballistic regime requires $a_{\rm
  u,d}<\gamma$, which, 
according to (\ref{weibelstrength}), is
fulfilled for unmagnetized pair shocks for all particles with
$\gamma>\gammabar$, provided $\ell_{\rm w}\lesssim10$, and $\sigma_{\rm u,d}<1$.
For electron-ion shocks, this condition is unlikely to be satisfied for
electrons with $\gamma\sim\bar{\gamma}$, 
but is easily satisfied for electrons that achieve
energies comparable to those of thermal ions or higher,
provided $\sigma_{\rm u,d}\ll 1$. 

\subsection{Helical transport regime}

In the helical transport regime, energy losses are important
at all points along a trajectory, and, if 
Bohm diffusion operates,
the time taken to return to the shock front is approximately 
the gyro period. Under these conditions, 
the 
maximum Lorentz factor has been calculated by
\citet{achterbergetal01}:
\eqb
\gamma&<&\frac{3 m^2 c^3}{2 e^3B}
\label{gammaachterberg}
\eqe

At magnetized, relativistic shocks, particle acceleration 
by the first-order Fermi mechanism is less plausible,
since, at least for superluminal shocks, it relies
on strong cross-field diffusion \citep[e.g.,][]{baringsummerlin09}.
However, if the process does operate, particles can 
move out of the helical regime into the ballistic regime, as their
Lorentz factor increases. Because of this, 
it is convenient to 
define a critical strength parameter $a_{\rm crit}$
such that when $a=a_{\rm crit}$ the maximum Lorentz 
factor $\gamma_{\rm max}$
permitted by radiation losses is achieved just at the point
at which the transport changes character from helical to ballistic. 
If $a>a_{\rm crit}$, all particles remain in the the helical regime.
On the other hand, if $a<a_{\rm crit}$, 
particles of the maximum Lorentz factor undergo ballistic transport, 
but lower energy particles may be in the helical regime.  
Since the transition from helical to ballistic occurs at 
$\gamma=a$, the critical strength parameter in
the downstream region follows from (\ref{wglimitdown}):
\eqb
a_{\rm crit}&=&\left(\frac{3mc^2\lambda_{\rm d}}{2e^2}\right)^{1/3}
\eqe
Expressed in these terms, the maximum Lorentz factor
in the helical transport regime (\ref{gammaachterberg})
is $\gamma_{\rm max}=\sqrt{a_{\rm crit}^3/a}$. Typically,
$a_{\rm crit}\gg1$ in both unmagnetized and magnetized shocks:
inserting, for example, the length scale appropriate for Weibel-mediated
shocks (\ref{weibellength}) one finds
\eqb
a_{\rm crit}&=&
1.4\times10^6\ell_{\rm w}^{1/3}{\bar{\gamma}}^{1/6}\left(n/1\,\textrm{cm}^{3}\right)^{-1/6}
\label{acritestimate}
\eqe

In the helical regime, the first-order Fermi process has been 
observed, so far, only 
in PIC simulations of subluminal pair shocks \citep{sironispitkovsky09a}.
The failure of the mechanism at superluminal shocks
may be because
the
helical character of 
orbits prevents 
particles from 
recrossing the shock
front by sweeping them away into the downstream medium
\citep{begelmankirk90,bednarzostrowski96,pelletier06}. 
However, first-order Fermi is not the only possible 
acceleration mechanism at superluminal shocks.
Strong electromagnetic 
wave fields are generated at these shocks, and, at least in
plasmas that contain some protons, these are thought to be
responsible for particle
acceleration \citep{amatoarons06,lyubarsky07}. 
This can only happen if the nonthermal
particles are confined within the source, i.e., if they are 
deflected through an angle $\sim 1$
before losing their energy to radiation. 
In this case, demanding that a particle be able to complete 
a gyration before radiating its energy leads to the 
the same constraint 
as obtained for first order Fermi acceleration
(\ref{gammaachterberg}).

Combining the constraints from the 
ballistic regime (\ref{wglimitdown}) and 
the helical regime (\ref{gammaachterberg}) gives:
\eqb
\gamma_{\rm max}&=&
\left\lbrace
\begin{array}{ll}
a_{\rm crit}&
\textrm{for\ }a<a_{\rm crit}\\
&\\
a_{\rm crit}\sqrt{a_{\rm crit}/a}&
\textrm{for\ }a>a_{\rm crit}
\end{array}
\right.
\label{sglimit}
\eqe

\section{Radiative signatures}

The spectrum of photons that are emitted when a particle is scattered
by turbulent fluctuations is, in the general case, 
quite complex \citep{toptyginetal87}.
However, to estimate the radiative signature of a
particle accelerated at a relativistic shock front several 
simplifications can be made. 

Consider the radiation emitted when a particle is scattered by a
single fluctuation of length $\lambda$ and 
strength parameter $a$, and is unperturbed
before and after interaction.  The character of the emission depends
crucially on the \lq\lq formation\rq\rq\ or \lq\lq coherence\rq\rq\
length of the radiated photons \cite{akhiezershulga87}.
At low frequencies, this length becomes large, but we confine our estimates to 
the highest frequency photons produced in the interaction. If 
$a>1$, the formation length is roughly $\lambda_{\rm coh}\approx
mc^2/eB$. This is 
much smaller than the wavelength of the turbulence,
so that the individual photons are created in regions in which the field is
almost constant and homogeneous. 
The result is synchrotron radiation, with the emissivity defined by the local
value of the field. The emission extends up to the roll-over frequency of the 
highest energy electrons:
\eqb
\omega_{\rm max}&\approx&0.5\gamma_{\rm max}^2 eB/mc
\nonumber\\
&=&0.5 a\gamma^2 c/\lambda\qquad\textrm{for\ }a>1
\label{omegamaxsynch}
\eqe
These photons are radiated into a forward directed cone of opening angle 
$1/\gamma$. Well below this frequency, the emissivity is 
proportional to $\omega^{1/3}$,
provided the coherence length remains shorter than the wavelength of the 
turbulence.

\begin{figure}
\includegraphics[bb=118 443 522 748,clip=true,width=0.5\textwidth]{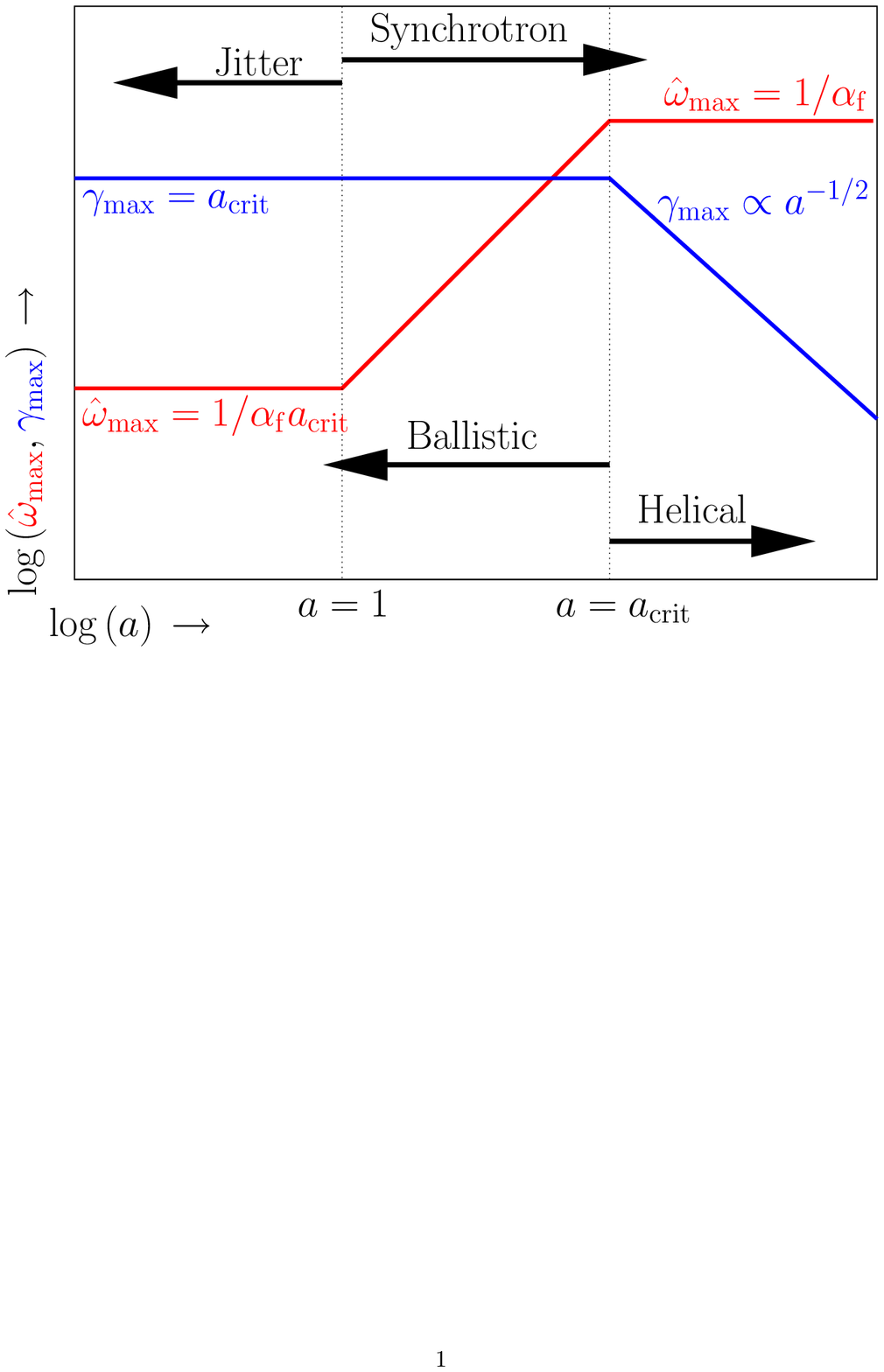}
\caption{The maximum Lorentz factor $\gamma_{\rm max}$ of electrons
and the maximum photon energy 
$\hat{\omega}_{\rm max}=\hbar\omega_{\rm max}/mc^2$
they radiate when 
scattered by magnetic fluctuations of 
strength $a$ (see Eqs.~(\ref{weibelstrength}) and 
(\ref{synchstrength})) at a relativistic shock. The critical strength
parameter $a_{\rm crit}$ is defined in (\ref{acritestimate}),
$\alpha_{\rm f}$ is the fine-structure constant. The jitter/synchrotron
regimes are separated by the vertical $a=1$ line; the ballistic/helical
transport regimes by the $a=a_{\rm crit}$ line.
See the electronic edition of the Journal for a color version 
of this figure.\label{sketch}}
\end{figure}

If, on the other hand, $a<1$, the particle is deflected through an
angle that is small compared to $1/\gamma$. 
In this case, the coherence length 
is no longer limited by deflection, but is given by
the distance moved by the particle
in the lab.\ frame during the time it takes for 
the photon to move one
wavelength ahead of the particle:
$\lambda_{\rm coh}\approx \gamma^2c/\omega$. 
The spectrum of radiated photons rolls over when 
$\lambda_{\rm coh}\approx\lambda$, and is flat ($\propto \omega^0$) 
towards lower frequency, as in the case of bremsstrahlung.
The maximum frequency is roughly
\eqb
\omega_{\rm max}&\approx&0.5\gamma_{\rm max}^2 c/\lambda
\qquad\textrm{for\ }a< 1
\label{omegamaxjitter}
\eqe

In each case, most of the power radiated by an individual electron
emerges within a decade of the roll-over frequency that corresponds to
its Lorentz factor. Therefore, to estimate the spectrum radiated by a power-law 
distribution of electrons,
with differential number density $\diff n/\diff\gamma\propto \gamma^{-p}$, 
one can simply make a one-to-one correspondence 
between radiated frequency and Lorentz factor: 
$\omega= \textrm{Max}(a,1)\gamma^2c/\lambda$. 
Larmor's formula, which holds for all values of $a$, states that 
the radiated power is proportional to the Lorentz factor squared,
so that the power radiated by $\diff n$ electrons of Lorentz factor $\gamma$ 
is $\diff L\propto\gamma^2\diff n$. A simple change of variables 
then shows that in both the synchrotron and jitter cases, the spectrum 
at frequencies below the roll-over frequency of the highest energy electrons
is $\diff L/\diff\omega\propto \omega^{-(p-1)/2}$. 
 
Combining the limit on the Lorentz factor
(\ref{wglimitdown}) with the expressions for the roll-over frequency
(\ref{omegamaxsynch}) and (\ref{omegamaxjitter}), one finds 
for the maximum frequency that can be radiated by particles accelerated 
at a relativistic shock front:
\eqb
\frac{\hbar\omega_{\rm max}}{mc^2}&=&
\left\lbrace
\begin{array}{ll}
\left(\alpha_{\rm f}a_{\rm crit}\right)^{-1}&a< 1
\\
&\\
a\left(\alpha_{\rm f}a_{\rm crit}\right)^{-1}&
1<a<a_{\rm crit}
\\
&\\
\alpha_{\rm f}^{-1}&a>a_{\rm crit}
\end{array}
\right.
\label{freqconstraint}
\eqe
where $\alpha_{\rm f}=e^2/\hbar c$ is the fine structure constant.
There are two transport regimes:
that in which the unperturbed motion of the highest energy particles 
is ballistic ($a<\gamma_{\rm max}$)
and that in which it is helical ($a>\gamma_{\rm max}$) 
and two radiation regimes: those of jitter
($a<1$) and synchrotron ($a>1$) radiation.
This is illustrated in Fig.~\ref{sketch}.

\section{Discussion}

Radiation emitted by relativistic electrons scattering in the
small-scale turbulent magnetic fields generated at Weibel-mediated
relativistic shocks has been proposed as the mechanism responsible for
both the prompt and afterglow emission of gamma-ray bursts
\citep{medvedev06,medvedevetal07}.  The evidence in favour of this
suggestion is based on modelling the observed spectra assuming an
electron distribution of power-law type with arbitrary high and low
energy cut-offs. Power-law distributions are expected on theoretical
grounds \citep{achterbergetal01,kirketal00}, and are indeed observed
in simulations of weakly magnetized, relativistic shocks
\citep{spitkovsky08b,sironispitkovsky09a,martinsetal09}.  However, the
constraint on the maximum photon energy imposed by the above analysis
(\ref{freqconstraint}) suggests that this picture is not
self-consistent, because the scatterings are too weak to accelerate
electrons to the required Lorentz factor.  
In order to radiate photons
of energy $\sim mc^2$ in the plasma rest frame, strong 
fluctuations of large length-scale 
with $a\sim\alpha_{\rm f}a_{\rm crit}\sim 10^4$ are
required.

The above conclusion rests on the assumption that the same fluctuations are
responsible for both the particle transport and radiation.
In terms of
the strength parameter $a$ and length scale $\lambda$ that we use to
characterize the fluctuations, the deflection angle scales as
$\Delta\theta\propto a$ and the radiation losses as
$\Delta\gamma\propto a^2/\lambda$. If, therefore, the scattering
responsible for isotropization occurs on fluctuations of comparable
strength, but much larger length scale than those responsible for the
radiation losses, the limit on the maximum photon energy
(\ref{freqconstraint}) is relaxed. In principle, the fluctuations induced
by the Weibel instability could be responsible for photon production,
provided longer wavelength fluctuations are present to provide the
necessary isotropization and transport. 
The accelerated particles themselves appear to 
generate longer wavelength fluctuations
downstream of the shock
\citep{keshetetal09}, but this is a relatively small effect
compared to that needed to significantly influence the maximum photon energy.

On the other hand, if, as simulations suggest, the Weibel-induced
fluctuations are responsible for the transport, the bulk of the
radiation must be produced by interaction with fluctuations of much
shorter wavelength. An obvious candidate is the soft photon field
produced by the interaction of thermal electrons with the
Weibel-induced fluctuations --- the \lq jitter\rq\ 
analog of the synchrotron photons produced by
relativistic thermal electrons
\citep{baringbraby04,gianniosspitkovsky09}.  With these photons as
targets, the radiation mechanism is analogous to the synchrotron
self-Compton mechanism, which has been discussed
in connection with the problem of rapidly
decaying magnetic fluctuations \citep{rossirees03}.  
While this model
offers a simple explanation for the production of high energy photons,
it does not easily accommodate
those
low-frequency BATSE burst spectra \citep{baringbraby04}, that 
violate the so-called synchrotron \lq line of
death\rq~\citep{preeceetal98}.  
These rare cases might be accounted
  for if the upscattering occurs deep in the Klein-Nishina regime
  \citep{derishevetal01,bosnjaketal09}, an explanation which 
is facilitated if the target photons arise not from synchrotron emission, but 
from relatively hard jitter radiation.

\acknowledgments This research was supported in part by the National
Science Foundation under Grant No.~PHY05-51164. BR gratefully
acknowledges support from the Alexander von Humboldt foundation.


\end{document}